# Recursive Total Least-Squares Algorithm Based on Inverse Power Method and Dichotomous Coordinate-Descent Iterations

Reza Arablouei, *Member, IEEE*, Kutluyıl Doğançay, *Senior Member, IEEE*, and Stefan Werner, *Senior Member, IEEE*

*Abstract*—We develop a recursive total least-squares (RTLS) algorithm for errors-in-variables system identification utilizing the inverse power method and the dichotomous coordinate-descent (DCD) iterations. The proposed algorithm, called DCD-RTLS, outperforms the previously-proposed RTLS algorithms, which are based on the line-search method, with reduced computational complexity. We perform a comprehensive analysis of the DCD-RTLS algorithm and show that it is asymptotically unbiased as well as being stable in the mean. We also find a lower bound for the forgetting factor that ensures mean-square stability of the algorithm and calculate the theoretical steady-state mean-square deviation (MSD). We verify the effectiveness of the proposed algorithm and the accuracy of the predicted steady-state MSD via simulations.

## I. INTRODUCTION

Most linear systems in signal processing, control, or econometrics applications can be described using the errors-in-variables (EIV) models. In the EIV models, both input and output data of a linear system are assumed to be contaminated with additive noise [1], [2]. The total least-squares (TLS) method is well suited for identification of the EIV models [3], [4]. Unlike the least-squares (LS) method that minimizes the sum of squared estimation errors without accounting for the noise in the input data, TLS is devised to minimize the perturbation in both input and output data that makes the input-output data fit through a linear system. Therefore, TLS is expected to have superior performance to LS when the input data as well as the output data is noisy.

It has been shown that the TLS estimate of a linear system is in fact the eigenvector corresponding to the eigenvalue of its augmented and weighted data covariance matrix that has the smallest absolute value [5]. A popular methodology to compute this eigenvector, also known as the *minor component*, is to minimize a Rayleigh quotient cost function [6] using any optimization technique such as gradient descent [7]-[9] or line search [10], [11]. Another common practice is to utilize the inverse power method (inverse iteration) [12]-[15].

The recursive TLS (RTLS) algorithms proposed in [10], [11], and [14] for adaptive finite-impulse-response (FIR) filtering have computational complexities of order $\mathcal{O}(L)$, where $L$ is the system order, when the input data is shift-structured. All three algorithms essentially accomplish the same task of computing the adaptive FIR filter weights by recursively estimating the minor component from the data available up to the current time. As a result, they perform very similarly at the steady state.

In [14], it is shown that the RTLS algorithm proposed therein, which is based on the inverse power method and a line-search strategy, is asymptotically consistent, i.e., converges to the optimal solution with probability one, when the perfect statistical knowledge of the input signal and noise is available or alternatively the forgetting factor is set to unity.

In this paper, we propose an RTLS algorithm for estimating the parameters of an EIV model for adaptive FIR filtering by employing the inverse power method together with the dichotomous coordinate-descent (DCD) iterations. We utilize the DCD iterations to solve two systems of linear equations associated with the calculation of two auxiliary vector variables. The DCD algorithm is a shift-and-add algorithm and can solve a system of linear equation using only additions and bit-shifts with no multiplication [16]-[18]. As a result, the proposed algorithm, called DCD-RTLS, has a computational complexity of $\mathcal{O}(L)$ when there is a shift structure in the input vectors. It requires $10L + 2$ multiplications as opposed to $15L + 11$, $16L + 19$, and $22L + 93$ multiplications required by the algorithms of [10], [11], and [14], respectively. In addition, simulation results show that the DCD-RTLS algorithm outperforms these algorithms in terms of convergence speed.

We examine the mean and mean-square performance of the DCD-RTLS algorithm under the assumption that the DCD algorithm can be made adequately accurate. We show that the DCD-RTLS algorithm is convergent in the mean and asymptotically unbiased as well as that, at the steady state, it

This work was supported in part by the Academy of Finland.
R. Arablouei and K. Doğançay are with the School of Engineering and the Institute for Telecommunications Research, University of South Australia, Mawson Lakes SA 5095, Australia (email: arary003@mymail.unisa.edu.au; kutluyil.dogancay@unisa.edu.au).
S. Werner is with the Department of Signal Processing and Acoustics, School of Electrical Engineering, Aalto University, Espoo, Finland (email: stefan.werner@aalto.fi).

performs akin to a bias-compensated recursive LS (BCRLS) algorithm. We find a lower bound for the forgetting factor that guarantees the mean-square stability of the algorithm. We also calculate the theoretical steady-state mean-square deviation (MSD) of the DCD-RTLS algorithm and corroborate its accuracy through numerical examples.

## II. SIGNAL AND SYSTEM MODEL

Let us consider a linear system of a-priori-known order $L \in \mathbb{N}$ described by

$$y_n = \mathbf{x}_n^T \mathbf{h} \tag{1}$$

where $\mathbf{h} \in \mathbb{R}^{L \times 1}$ is the column vector of the system parameters, superscript $T$ denotes matrix transposition, and $\mathbf{x}_n \in \mathbb{R}^{L \times 1}$ and $y_n \in \mathbb{R}$ are the input and output of the system at time index $n \in \mathbb{N}$, respectively. We observe noisy versions of $\mathbf{x}_n$ and $y_n$, i.e.,

$$\tilde{\mathbf{x}}_n = \mathbf{x}_n + \mathbf{u}_n \tag{2}$$

and

$$\tilde{y}_n = y_n + v_n \tag{3}$$

where $\mathbf{u}_n \in \mathbb{R}^{L \times 1}$ and $v_n \in \mathbb{R}$ are the corresponding input and output noises. Substituting (2) and (3) into (1) gives

$$\tilde{y}_n = \mathbf{h}^T(\tilde{\mathbf{x}}_n - \mathbf{u}_n) + v_n. \tag{4}$$

We make the following assumptions regarding the noises:

A1: The entries of the input noise vector are wide-sense stationary with zero mean and variance $\eta \in \mathbb{R}_{\geq 0}$.

A2: The output noise is wide-sense stationary with zero mean and variance $\xi \in \mathbb{R}_{\geq 0}$.

We also define

$$\tilde{\mathbf{X}}_n = \left[ \lambda^{(n-1)/2} \tilde{\mathbf{x}}_1, \lambda^{(n-2)/2} \tilde{\mathbf{x}}_2, \dots, \lambda^{1/2} \tilde{\mathbf{x}}_{n-1}, \tilde{\mathbf{x}}_n \right],$$

$$\tilde{\mathbf{y}}_n = \left[ \lambda^{(n-1)/2} \tilde{y}_1, \lambda^{(n-2)/2} \tilde{y}_2, \dots, \lambda^{1/2} \tilde{y}_{n-1}, \tilde{y}_n \right],$$

$$\begin{aligned}
\boldsymbol{\Phi}_n &= \tilde{\mathbf{X}}_n \tilde{\mathbf{X}}_n^T \\
&= \lambda \boldsymbol{\Phi}_{n-1} + \tilde{\mathbf{x}}_n \tilde{\mathbf{x}}_n^T,
\end{aligned} \tag{5}$$

$$\begin{aligned}
\mathbf{z}_n &= \tilde{\mathbf{X}}_n \tilde{\mathbf{y}}_n^T \\
&= \lambda \mathbf{z}_{n-1} + \tilde{y}_n \tilde{\mathbf{x}}_n
\end{aligned} \tag{6}$$

$$\begin{aligned}
\tau_n &= \tilde{\mathbf{y}}_n \tilde{\mathbf{y}}_n^T \\
&= \lambda \tau_{n-1} + \tilde{y}_n^2,
\end{aligned}$$

$$\gamma = \xi / \eta,$$

$$\mathbf{D} = \begin{bmatrix} \mathbf{I} & \mathbf{0} \\ \mathbf{0}^T & \gamma^{-1/2} \end{bmatrix},$$

and

$$\begin{aligned}
\boldsymbol{\Psi}_n &= \mathbf{D}^T \begin{bmatrix} \tilde{\mathbf{X}}_n \\ \tilde{\mathbf{y}}_n \end{bmatrix} [\tilde{\mathbf{X}}_n^T, \tilde{\mathbf{y}}_n^T] \mathbf{D} \\
&= \begin{bmatrix} \boldsymbol{\Phi}_n & \gamma^{-1/2} \mathbf{z}_n \\ \gamma^{-1/2} \mathbf{z}_n^T & \gamma^{-1} \tau_n \end{bmatrix}
\end{aligned}$$

where $\lambda \in \mathbb{R}_{>0}$ is a forgetting factor that satisfies

$$0 \ll \lambda < 1,$$

$\mathbf{0}$ is the $L \times 1$ zero vector, and $\mathbf{I}$ the $L \times L$ identity matrix.

## III. ALGORITHM DESCRIPTION

### A. Recursive Total Least-Squares

The TLS estimate of the system parameters at time instant $n$, denoted by $\mathbf{w}_n \in \mathbb{R}^{L \times 1}$, is given by

$$\begin{bmatrix} \mathbf{w}_n \\ -1 \end{bmatrix} = -\frac{\mathbf{q}_n}{\gamma^{-1/2} q_{L+1,n}}$$

where $\mathbf{q}_n \in \mathbb{R}^{(L+1) \times 1}$ is the eigenvector corresponding to the smallest (in absolute value) eigenvalue of the augmented and weighted data covariance matrix $\boldsymbol{\Psi}_n$ and $q_{L+1,n}$ is $(L+1)$th entry of $\mathbf{q}_n$ [5].

The eigenvector $\mathbf{q}_n$ can be estimated adaptively by executing a single iteration of the inverse power method [19] at each time instant using the following recursion:

$$\mathbf{q}_n = \boldsymbol{\Psi}_n^{-1} \mathbf{q}_{n-1}. \tag{7}$$

Note that the convergence rate of the inverse power method depends on the ratio of the two smallest eigenvalues of the matrix $\boldsymbol{\Psi}_n$. If these two eigenvalues are very close, the inverse power method will converge slowly or may even fail to converge.

Multiplying both sides of (7) by $\boldsymbol{\Psi}_n / (\gamma^{-1/2} q_{L+1,n-1} q_{L+1,n})$ gives

$$\boldsymbol{\Psi}_n \begin{bmatrix} \mathbf{w}_n \\ -\gamma^{1/2} \end{bmatrix} = \frac{q_{L+1,n-1}}{q_{L+1,n}} \begin{bmatrix} \mathbf{w}_{n-1} \\ -\gamma^{1/2} \end{bmatrix}$$

or equivalently

$$\boldsymbol{\Phi}_n \mathbf{w}_n - \mathbf{z}_n = \frac{q_{L+1,n-1}}{q_{L+1,n}} \mathbf{w}_{n-1} \tag{8}$$

and

$$\mathbf{z}_n^T \mathbf{w}_n - \tau_n = -\gamma \frac{q_{L+1,n-1}}{q_{L+1,n}}. \tag{9}$$

Substituting (9) into (8) gives

$$\boldsymbol{\Phi}_n \mathbf{w}_n - \mathbf{z}_n = -\gamma^{-1} \mathbf{w}_{n-1} \mathbf{z}_n^T \mathbf{w}_n + \gamma^{-1} \tau_n \mathbf{w}_{n-1}. \tag{10}$$

Solving (10) for $\mathbf{w}_n$ yields an RTLS estimate of $\mathbf{h}$ as

$$\mathbf{w}_n = (\boldsymbol{\Phi}_n + \gamma^{-1} \mathbf{w}_{n-1} \mathbf{z}_n^T)^{-1} (\mathbf{z}_n + \gamma^{-1} \tau_n \mathbf{w}_{n-1}). \tag{11}$$

Employing the Sherman-Morrison formula [19], we can avoid the matrix inversion in (11) by writing it as

$$\mathbf{w}_n = \left( \boldsymbol{\Phi}_n^{-1} - \frac{\boldsymbol{\Phi}_n^{-1} \mathbf{w}_{n-1} \mathbf{z}_n^T \boldsymbol{\Phi}_n^{-1}}{\gamma + \mathbf{z}_n^T \boldsymbol{\Phi}_n^{-1} \mathbf{w}_{n-1}} \right) (\mathbf{z}_n + \gamma^{-1} \tau_n \mathbf{w}_{n-1}) \tag{12}$$

where $\boldsymbol{\Phi}_n^{-1}$ can be updated via

$$\boldsymbol{\Phi}_n^{-1} = \lambda^{-1} \boldsymbol{\Phi}_{n-1}^{-1} - \frac{\lambda^{-1} \boldsymbol{\Phi}_{n-1}^{-1} \tilde{\mathbf{x}}_n \tilde{\mathbf{x}}_n^T \boldsymbol{\Phi}_{n-1}^{-1}}{\lambda + \tilde{\mathbf{x}}_n^T \boldsymbol{\Phi}_{n-1}^{-1} \tilde{\mathbf{x}}_n}.$$

The computational complexities of (11) and (12) are $\mathcal{O}(L^3)$ and $\mathcal{O}(L^2)$, respectively, even with shift-structured input data.

## B. Utilization of the DCD Algorithm

Defining

$$\mathbf{m}_{1,n} = \boldsymbol{\Phi}_n^{-1}\mathbf{z}_n \qquad (13)$$

and

$$\mathbf{m}_{2,n} = \boldsymbol{\Phi}_n^{-1}\mathbf{w}_{n-1}, \qquad (14)$$

(12) can be written as

$$\mathbf{w}_n = \mathbf{m}_{1,n} + \gamma^{-1}\tau_n \mathbf{m}_{2,n} - \frac{\mathbf{z}_n^T(\mathbf{m}_{1,n} + \gamma^{-1}\tau_n \mathbf{m}_{2,n})}{\gamma + \mathbf{z}_n^T \mathbf{m}_{2,n}}\mathbf{m}_{2,n}. \qquad (15)$$

In order to reduce the computational complexity, instead of calculating $\mathbf{m}_{1,n}$ and $\mathbf{m}_{2,n}$ directly from (13) and (14), we can compute them by solving the following systems of linear equations utilizing the DCD iterations and exerting only additions and bit-shifts:

$$\boldsymbol{\Phi}_n \mathbf{m}_{1,n} = \mathbf{z}_n \qquad (16)$$

and

$$\boldsymbol{\Phi}_n \mathbf{m}_{2,n} = \mathbf{w}_{n-1}. \qquad (17)$$

Furthermore, to exploit the full potential of the DCD algorithm and to minimize the number of its required iterations, we can rewrite (16) and (17) as

$$\boldsymbol{\Phi}_n \mathbf{d}_{1,n} = \mathbf{p}_{1,n} \qquad (18)$$

and

$$\boldsymbol{\Phi}_n \mathbf{d}_{2,n} = \mathbf{p}_{2,n} \qquad (19)$$

where

$$\mathbf{d}_{1,n} = \mathbf{m}_{1,n} - \mathbf{m}_{1,n-1},$$

$$\mathbf{d}_{2,n} = \mathbf{m}_{2,n} - \mathbf{m}_{2,n-1},$$

$$\mathbf{p}_{1,n} = \mathbf{z}_n - \boldsymbol{\Phi}_n \mathbf{m}_{1,n-1}, \qquad (20)$$

and

$$\mathbf{p}_{2,n} = \mathbf{w}_{n-1} - \boldsymbol{\Phi}_n \mathbf{m}_{2,n-1} \qquad (21)$$

and solve (18) and (19) instead of (16) and (17).

The precision of the solutions to the systems of linear equations in (18) and (19) provided by the DCD algorithm can be represented by the residual vectors that are defined as

$$\mathbf{r}_{1,n-1} = \mathbf{z}_{n-1} - \boldsymbol{\Phi}_{n-1}\mathbf{m}_{1,n-1} \qquad (22)$$

and

$$\mathbf{r}_{2,n-1} = \mathbf{w}_{n-2} - \boldsymbol{\Phi}_{n-1}\mathbf{m}_{2,n-1} \qquad (23)$$

at time instant $n-1$. We can use these residual vectors to improve the accuracy of the solutions at time instant $n$. Substitution of (5) and (6) into (20) and (21) together with using (22) and (23) results in

$$\mathbf{p}_{1,n} = \lambda \mathbf{r}_{1,n-1} + \left(\tilde{y}_n - \tilde{\mathbf{x}}_n^T \mathbf{m}_{1,n-1}\right)\tilde{\mathbf{x}}_n$$

and

$$\mathbf{p}_{2,n} = \lambda \mathbf{r}_{2,n-1} + \mathbf{w}_{n-1} - \lambda \mathbf{w}_{n-2} - \left(\tilde{\mathbf{x}}_n^T \mathbf{m}_{2,n-1}\right)\tilde{\mathbf{x}}_n.$$

Solving (18) and (19) using the DCD algorithm yields $\mathbf{d}_{1,n}$, $\mathbf{d}_{2,n}$, $\mathbf{r}_{1,n}$, and $\mathbf{r}_{2,n}$ at each time instant $n$ [17]. Having calculated $\mathbf{d}_{1,n}$ and $\mathbf{d}_{2,n}$, we obtain $\mathbf{m}_{1,n}$ and $\mathbf{m}_{2,n}$ using

$$\mathbf{m}_{1,n} = \mathbf{m}_{1,n-1} + \mathbf{d}_{1,n}$$

and

$$\mathbf{m}_{2,n} = \mathbf{m}_{2,n-1} + \mathbf{d}_{2,n}.$$

We can then update the filter weight vector $\mathbf{w}_n$, which is an estimate of $\mathbf{h}$ at time instant $n$, using (15).

We summarize the proposed DCD-RTLS algorithm in Table I. We present the DCD algorithm solving (18) in Table II where $r_{l,i,n}$ and $d_{l,i,n}$ denote the $l$th entry of the vectors $\mathbf{r}_{i,n}$ and $\mathbf{d}_{i,n}$, respectively, while $\phi_{l,l,n}$ and $\boldsymbol{\phi}_{l,n}$ are the $(l,l)$th entry and the $l$th column of the matrix $\boldsymbol{\Phi}_n$, respectively. In Tables I and II, we also give the number of arithmetic operations required by every step of the algorithms.

Three design parameters, $N \in \mathbb{N}$, $M \in \mathbb{N}$, and $H \in \mathbb{R}$ govern the accuracy and complexity of the DCD algorithm [16]. The parameter $M$ is the number of bits used to represent the entries of the solution vectors, $\mathbf{d}_{1,n}$ and $\mathbf{d}_{2,n}$, as fixed-point words within an amplitude range of $[-H, H]$. The parameters $M$ and $H$ are generally set based on empirical knowledge in conjunction with desired accuracy. The DCD algorithm renders maximum $N$ iterative updates at each run. Thus, $N$ plays a key role in determining the precision of the DCD algorithm. Typically, the larger $N$ is, the more accurate the result is. However, exploiting the information within the residual vector ($\mathbf{r}_{1,n-1}$ or $\mathbf{r}_{2,n-1}$ in our case) often dramatically decreases the number of DCD iterations required to achieve a satisfactory result. Furthermore, $N$ delimits the maximum number of entries in $\mathbf{m}_{1,n}$ and $\mathbf{m}_{2,n}$ that are updated at each time instant. Consequently, when $N < L$, the DCD iterations implements a form of *selective partial updates* [20], [21]. While the DCD algorithm affords tremendous savings in terms of computation complexity, the requirement of tuning three design parameters can be considered as its main disadvantage.

## IV. COMPUTATIONAL COMPLEXITY

Since $\boldsymbol{\Phi}_n$ is symmetric, it is sufficient to update only its upper-triangular part. Moreover, by selecting the forgetting factor as $\lambda = 1 - 2^{-P}$ where $P$ is a positive integer, we can replace multiplications by $\lambda$ with additions and bit-shifts [17], [18]. When there is shift structure in the input data, i.e., the input vector has the following form:

$$\tilde{\mathbf{x}}_n = [\tilde{x}_n, \tilde{x}_{n-1}, \ldots, \tilde{x}_{n-L+1}]^T$$

where $\tilde{x}_n \in \mathbb{R}$ is the noisy input signal, updating $\boldsymbol{\Phi}_n$ is significantly simplified. In this case, the upper-left $(L-1) \times (L-1)$ block of $\boldsymbol{\Phi}_{n-1}$ can be copied to the lower-right $(L-1) \times (L-1)$ block of $\boldsymbol{\Phi}_n$. Thus, only the first row and the first column of $\boldsymbol{\Phi}_n$ are directly updated. Due to the symmetry of $\boldsymbol{\Phi}_n$, it is sufficient to update only the first column.

In Table III, we present the number of required multiplication, addition, division, and square-root operations per iteration by the DCD-RTLS algorithm and the algorithms of [10], [11], and [14], which are called kRTLS, xRTLS, and AIP, respectively. In Fig. 1, we plot the total number of required gates for fixed-point hardware implementation of different algorithms versus the system order, $L$, when $M = 16$ and $N = 1$ in the DCD-RTLS algorithm. We assume a unit-gate area model where a 16-bit carry-lookahead adder requires 204 gates and a 16-bit array multiplier requires 2,336 gates [22]-[25]. For simplicity, we presume that a division or square-root operation requires the same number of gates as a multiplication operation. We consider both cases of shift-structured and non-shift-structured input data.

## V. Performance Analysis

Analyzing the performance of the DCD-RTLS algorithm taking into account the impreciseness sustained by the DCD algorithm is arduous. Therefore, we assume that the solutions provided by the DCD algorithm are sufficiently accurate so that we can neglect any error due to the approximate nature of the DCD algorithm. It has been shown that usually an appropriate choice of the design parameters, especially $N$, makes the DCD algorithm sufficiently accurate hence this assumption acceptable [17], [18], [25]-[28]. Thus, in this section, we provide a thorough performance analysis of the RTLS algorithm, i.e., the recursion of (11). The analysis results are applicable to the DCD-RTLS algorithm given that the design parameters of the DCD algorithm are properly chosen so that it solves (18) and (19) with sufficient accuracy.

### A. Assumptions

For tractability of the analysis, we adopt the following additional assumptions:

A3: The noises are temporally uncorrelated and statistically independent of each other as well as the noiseless input data.

A4: The noiseless input vector, $\mathbf{x}_n$, is wide-sense stationary and temporally uncorrelated, with a positive-definite covariance matrix $\mathbf{R} \in \mathbb{R}^{L \times L}$.

A5: For a sufficiently large $n$, we can replace $\boldsymbol{\Phi}_n$, $\mathbf{z}_n$, and $\tau_n$ with their asymptotic expected values, $\bar{\boldsymbol{\Phi}}_\infty$, $\bar{\mathbf{z}}_\infty$, and $\bar{\tau}_\infty$, respectively, which, considering A1-A4, are calculated as

$$\bar{\boldsymbol{\Phi}}_\infty = \lim_{n \to \infty} E[\boldsymbol{\Phi}_n] = (1 - \lambda)^{-1}(\mathbf{R} + \eta \mathbf{I}),$$

$$\bar{\mathbf{z}}_\infty = \lim_{n \to \infty} E[\mathbf{z}_n] = (1 - \lambda)^{-1} \mathbf{Rh},$$

and

$$\bar{\tau}_\infty = \lim_{n \to \infty} E[\tau_n] = (1 - \lambda)^{-1}(\mathbf{h}^T \mathbf{Rh} + \xi).$$

### B. Mean Convergence

Taking the expectation of both sides of (11) after a sufficiently large number of iterations while bearing in mind A5 and the Slutsky's theorem [29] results in

$$\begin{aligned} \bar{\mathbf{w}}_n &= (\bar{\boldsymbol{\Phi}}_\infty + \gamma^{-1} \bar{\mathbf{w}}_{n-1} \bar{\mathbf{z}}_\infty^T)^{-1}(\bar{\mathbf{z}}_\infty + \gamma^{-1} \bar{\tau}_\infty \bar{\mathbf{w}}_{n-1}) \\ &= (\mathbf{R} + \eta \mathbf{I} + \gamma^{-1} \bar{\mathbf{w}}_{n-1} \mathbf{h}^T \mathbf{R})^{-1} \\ &\quad \times [\mathbf{Rh} + (\gamma^{-1} \mathbf{h}^T \mathbf{Rh} + \eta) \bar{\mathbf{w}}_{n-1}] \end{aligned} \quad (24)$$

where, for convenience of the notation, we define

$$\bar{\mathbf{w}}_n = E[\mathbf{w}_n].$$

Using (24), we can verify that

$$\bar{\mathbf{w}}_n - \mathbf{h} = \eta (\mathbf{R} + \eta \mathbf{I} + \gamma^{-1} \bar{\mathbf{w}}_{n-1} \mathbf{h}^T \mathbf{R})^{-1} (\bar{\mathbf{w}}_{n-1} - \mathbf{h}) \quad (25)$$

and

$$\begin{aligned} \gamma^{-1} \bar{\mathbf{w}}_n \mathbf{h}^T + \mathbf{I} &= (\mathbf{R} + \eta \mathbf{I} + \gamma^{-1} \bar{\mathbf{w}}_{n-1} \mathbf{h}^T \mathbf{R})^{-1} \\ &\quad \times (\gamma^{-1} \bar{\mathbf{w}}_{n-1} \mathbf{h}^T + \mathbf{I})(\gamma^{-1} \mathbf{Rhh}^T + \mathbf{R} + \eta \mathbf{I}). \end{aligned} \quad (26)$$

Inverting both sides of (26) then multiplying by (25) from the right gives

$$\mathbf{a}_n = \mathbf{Ca}_{n-1}$$

where

$$\mathbf{a}_n = (\gamma^{-1} \bar{\mathbf{w}}_n \mathbf{h}^T + \mathbf{I})^{-1} (\bar{\mathbf{w}}_n - \mathbf{h})$$

and

$$\mathbf{C} = \eta (\gamma^{-1} \mathbf{Rhh}^T + \mathbf{R} + \eta \mathbf{I})^{-1}.$$

Since $\mathbf{R}$ and $\gamma^{-1} \mathbf{hh}^T + \mathbf{I}$ are symmetric positive-definite, their multiplication, $\gamma^{-1} \mathbf{Rhh}^T + \mathbf{R}$, is also symmetric positive-definite with all positive eigenvalues. Therefore, the spectral radius of the matrix $\mathbf{C}$ is equal to

$$\begin{aligned} \rho\{\mathbf{C}\} &= \eta \zeta_{\max}\{(\gamma^{-1} \mathbf{Rhh}^T + \mathbf{R} + \eta \mathbf{I})^{-1}\} \\ &= \frac{\eta}{\zeta_{\min}\{\gamma^{-1} \mathbf{Rhh}^T + \mathbf{R}\} + \eta} \end{aligned} \quad (27)$$

where $\zeta_{\max}\{\cdot\}$ and $\zeta_{\min}\{\cdot\}$ return the eigenvalues of their matrix arguments that have the largest and smallest absolute values, respectively. From (27), we observe that

$$\rho\{\mathbf{C}\} < 1$$

and consequently

$$\lim_{n \to \infty} \mathbf{a}_n = \mathbf{0}$$

or

$$\lim_{n \to \infty} E[\mathbf{w}_n] = \mathbf{h}.$$

This indicates that the RTLS algorithm is convergent in the mean and asymptotically unbiased.

### C. Bias Compensation Mechanism

Let us rearrange (10) as

$$\mathbf{w}_n = \boldsymbol{\Phi}_n^{-1} \mathbf{z}_n + \gamma^{-1} (\tau_n - \mathbf{z}_n^T \mathbf{w}_n) \boldsymbol{\Phi}_n^{-1} \mathbf{w}_{n-1} \quad (28)$$

and make the following approximation after a sufficiently large number of iterations:

$$\tau_n - \mathbf{z}_n^T \mathbf{w}_n \approx \bar{\tau}_\infty - \bar{\mathbf{z}}_\infty^T \mathbf{h}$$
$$\approx (1-\lambda)^{-1}\xi. \quad (29)$$

Substituting (29) into (28), gives

$$\mathbf{w}_n = \mathbf{\Phi}_n^{-1} \mathbf{z}_n + (1-\lambda)^{-1}\eta \mathbf{\Phi}_n^{-1} \mathbf{w}_{n-1}. \quad (30)$$

The first term on the right-hand side of (30) is the conventional exponentially-weighted recursive LS (RLS) estimate [30], [31]. It is known that, in the presence of input noise, the RLS estimate is biased. The asymptotic bias of the RLS estimate is calculated as

$$\mathbf{b} = \lim_{n \to \infty} E[\mathbf{\Phi}_n^{-1} \mathbf{z}_n - \mathbf{h}]$$
$$= \bar{\mathbf{\Phi}}_\infty^{-1} \bar{\mathbf{z}}_\infty - \mathbf{h}$$
$$= (\mathbf{R} + \eta \mathbf{I})^{-1} \mathbf{R}\mathbf{h} - \mathbf{h}$$

and using the Woodbury matrix identity [19] as

$$\mathbf{b} = -\eta(\mathbf{R} + \eta \mathbf{I})^{-1}\mathbf{h}.$$

Approximating $(\mathbf{R} + \eta \mathbf{I})^{-1}$ with $(1-\lambda)^{-1}\mathbf{\Phi}_n^{-1}$ and $\mathbf{h}$ with $\mathbf{w}_{n-1}$, when $n$ is sufficiently large, gives

$$\mathbf{b} \approx -(1-\lambda)^{-1}\eta \mathbf{\Phi}_n^{-1} \mathbf{w}_{n-1}. \quad (31)$$

In view of (31), (30) resembles the update equation of a BCRLS algorithm [32]-[34] demonstrating how the RTLS algorithm removes the estimation bias from the RLS estimate. Moreover, since (30) is more tractable than the original RTLS recursion, (11), we will utilize it as the steady-state approximation of the RTLS algorithm in our later analysis.

### D. Weight-Error Update Equation

Multiplying both sides of (30) by $\mathbf{\Phi}_n$ from the left then subtracting $(1-\lambda)^{-1}\eta \mathbf{w}_n$ from it gives

$$\acute{\mathbf{\Phi}}_n \mathbf{w}_n = \mathbf{z}_n + (1-\lambda)^{-1}\eta(\mathbf{w}_{n-1} - \mathbf{w}_n) \quad (32)$$

where we define

$$\acute{\mathbf{\Phi}}_n = \mathbf{\Phi}_n - (1-\lambda)^{-1}\eta \mathbf{I}.$$

At time instant $n-1$, (32) is written as

$$\acute{\mathbf{\Phi}}_{n-1} \mathbf{w}_{n-1} = \mathbf{z}_{n-1} + (1-\lambda)^{-1}\eta(\mathbf{w}_{n-2} - \mathbf{w}_{n-1}). \quad (33)$$

Multiplying both sides of (33) by $\lambda$ and using the recursive equations

$$\mathbf{z}_n = \lambda \mathbf{z}_{n-1} + \tilde{y}_n \tilde{\mathbf{x}}_n$$

and

$$\mathbf{\Phi}_n = \lambda \mathbf{\Phi}_{n-1} + \tilde{\mathbf{x}}_n \tilde{\mathbf{x}}_n^T,$$

we get

$$\acute{\mathbf{\Phi}}_n \mathbf{w}_{n-1} - \tilde{\mathbf{x}}_n \tilde{\mathbf{x}}_n^T \mathbf{w}_{n-1} + \eta \mathbf{w}_{n-1} = \mathbf{z}_n - \tilde{y}_n \tilde{\mathbf{x}}_n$$
$$+ \lambda(1-\lambda)^{-1}\eta(\mathbf{w}_{n-2} - \mathbf{w}_{n-1}). \quad (34)$$

Subtracting (34) from (32) gives

$$\acute{\mathbf{\Phi}}_n \mathbf{w}_n = \acute{\mathbf{\Phi}}_n \mathbf{w}_{n-1} - \tilde{\mathbf{x}}_n \tilde{\mathbf{x}}_n^T \mathbf{w}_{n-1} + \eta \mathbf{w}_{n-1} + \tilde{y}_n \tilde{\mathbf{x}}_n$$
$$+ (1-\lambda)^{-1}\eta(\mathbf{w}_{n-1} - \mathbf{w}_n + \lambda \mathbf{w}_{n-1} - \lambda \mathbf{w}_{n-2}). \quad (35)$$

Subtracting both sides of (35) from $\acute{\mathbf{\Phi}}_n \mathbf{h}$ along with using (4) and assuming that, for a large $n$, the last term on the right-hand side of (35) is negligible results in

$$\acute{\mathbf{\Phi}}_n \breve{\mathbf{w}}_n = \acute{\mathbf{\Phi}}_n \breve{\mathbf{w}}_{n-1} - (\tilde{\mathbf{x}}_n \tilde{\mathbf{x}}_n^T - \eta \mathbf{I})\breve{\mathbf{w}}_{n-1}$$
$$+ (\tilde{\mathbf{x}}_n \mathbf{u}_n^T - \eta \mathbf{I})\mathbf{h} - \nu_n \tilde{\mathbf{x}}_n. \quad (36)$$

where we define the weight-error vector as

$$\breve{\mathbf{w}}_n = \mathbf{h} - \mathbf{w}_n.$$

Owing to A5, for a sufficiently large $n$, we have

$$\acute{\mathbf{\Phi}}_n \approx \bar{\mathbf{\Phi}}_\infty - (1-\lambda)^{-1}\eta \mathbf{I}$$
$$\approx (1-\lambda)^{-1}\mathbf{R}. \quad (37)$$

Substituting (37) into (36) and multiplying both sides by $(1-\lambda)\mathbf{R}^{-1}$ gives an approximate weight-error update equation for the RTLS algorithm as

$$\breve{\mathbf{w}}_n = [\mathbf{I} - (1-\lambda)\mathbf{R}^{-1}(\tilde{\mathbf{x}}_n \tilde{\mathbf{x}}_n^T - \eta \mathbf{I})]\breve{\mathbf{w}}_{n-1}$$
$$+ (1-\lambda)\mathbf{R}^{-1}[(\tilde{\mathbf{x}}_n \mathbf{u}_n^T - \eta \mathbf{I})\mathbf{h} - \nu_n \tilde{\mathbf{x}}_n]. \quad (38)$$

### E. Mean-Square Convergence

The assumptions A1-A4 imply the following corollary:

C1: The vector $\breve{\mathbf{w}}_{n-1}$ is statistically independent of $\tilde{\mathbf{x}}_n \tilde{\mathbf{x}}_n^T$, $\tilde{\mathbf{x}}_n \mathbf{u}_n^T$, and $\nu_n \tilde{\mathbf{x}}_n$.

Taking the expectation of the squared Euclidean norm of both sides of (38) while considering C1 gives the following variance relation:

$$E[\|\breve{\mathbf{w}}_n\|^2] = E[\breve{\mathbf{w}}_{n-1}^T \mathbf{S}_n \breve{\mathbf{w}}_{n-1}] + (1-\lambda)^2 g \quad (39)$$

where

$$\mathbf{S}_n = [\mathbf{I} - (1-\lambda)(\tilde{\mathbf{x}}_n \tilde{\mathbf{x}}_n^T - \eta \mathbf{I})\mathbf{R}^{-1}]$$
$$\times [\mathbf{I} - (1-\lambda)\mathbf{R}^{-1}(\tilde{\mathbf{x}}_n \tilde{\mathbf{x}}_n^T - \eta \mathbf{I})],$$

$$g = E[\mathbf{t}_n^T \mathbf{R}^{-2} \mathbf{t}_n]$$
$$= \text{tr}\{\mathbf{R}^{-2} E[\mathbf{t}_n \mathbf{t}_n^T]\}, \quad (40)$$

and

$$\mathbf{t}_n = (\tilde{\mathbf{x}}_n \mathbf{u}_n^T - \eta \mathbf{I})\mathbf{h} - \nu_n \tilde{\mathbf{x}}_n.$$

In view of A1-A4, we have

$$E[\mathbf{t}_n \mathbf{t}_n^T] = E[(\tilde{\mathbf{x}}_n \mathbf{u}_n^T - \eta \mathbf{I})\mathbf{h}\mathbf{h}^T(\mathbf{u}_n \tilde{\mathbf{x}}_n^T - \eta \mathbf{I})] + E[\nu_n^2 \tilde{\mathbf{x}}_n \tilde{\mathbf{x}}_n^T]$$
$$= E[\tilde{\mathbf{x}}_n \mathbf{u}_n^T \mathbf{h}\mathbf{h}^T \mathbf{u}_n \tilde{\mathbf{x}}_n^T] + \eta^2 \mathbf{h}\mathbf{h}^T - \eta E[\tilde{\mathbf{x}}_n \mathbf{u}_n^T]\mathbf{h}\mathbf{h}^T$$
$$- \eta \mathbf{h}\mathbf{h}^T E[\mathbf{u}_n \tilde{\mathbf{x}}_n^T] + E[\nu_n^2 \tilde{\mathbf{x}}_n \tilde{\mathbf{x}}_n^T]$$
$$= E[\tilde{\mathbf{x}}_n \mathbf{u}_n^T \mathbf{h}\mathbf{h}^T \mathbf{u}_n \tilde{\mathbf{x}}_n^T] - \eta^2 \mathbf{h}\mathbf{h}^T + \xi(\mathbf{R} + \eta \mathbf{I}). \quad (41)$$

Utilizing the Isserlis' theorem [35], we also have

$$E[\tilde{\mathbf{x}}_n \mathbf{u}_n^T \mathbf{h}\mathbf{h}^T \mathbf{u}_n \tilde{\mathbf{x}}_n^T] = E[\mathbf{h}^T \mathbf{u}_n \mathbf{u}_n^T \mathbf{h}]E[\tilde{\mathbf{x}}_n \tilde{\mathbf{x}}_n^T]$$
$$+ 2E[\tilde{\mathbf{x}}_n \mathbf{u}_n^T \mathbf{h}]E[\mathbf{h}^T \mathbf{u}_n \tilde{\mathbf{x}}_n^T] \quad (42)$$
$$= \eta \|\mathbf{h}\|^2 (\mathbf{R} + \eta \mathbf{I}) + 2\eta^2 \mathbf{h}\mathbf{h}^T,$$

Substituting (42) into (41) and then the resulting expression into (40) yields

$$g = \text{tr}\{\mathbf{R}^{-2}[(\eta \|\mathbf{h}\|^2 + \xi)(\mathbf{R} + \eta \mathbf{I}) + \eta^2 \mathbf{h}\mathbf{h}^T]\}. \quad (43)$$

In view of C1, $\check{\mathbf{w}}_{n-1}$ is independent of $\mathbf{S}_n$. Hence, we can write

$$E[\check{\mathbf{w}}_{n-1}^T \mathbf{S}_n \check{\mathbf{w}}_{n-1}] = E[\check{\mathbf{w}}_{n-1}^T \bar{\mathbf{S}} \check{\mathbf{w}}_{n-1}].$$

where

$$\begin{aligned}\bar{\mathbf{S}} &= E[\mathbf{S}_n] \\ &= \mathbf{I} - 2(1-\lambda)\mathbf{I} \\ &\quad + (1-\lambda)^2 E[(\tilde{\mathbf{x}}_n \tilde{\mathbf{x}}_n^T - \eta \mathbf{I})\mathbf{R}^{-2}(\tilde{\mathbf{x}}_n \tilde{\mathbf{x}}_n^T - \eta \mathbf{I})] \\ &= (2\lambda - 1)\mathbf{I} \\ &\quad + (1-\lambda)^2 (\text{tr}\{\mathbf{R}^{-1}\}\mathbf{R} + 2\mathbf{I} - 2\eta \mathbf{R}^{-1} + \eta^2 \mathbf{R}^{-2}) \\ &= (1 - 2\lambda + 2\lambda^2)\mathbf{I} \\ &\quad + (1-\lambda)^2 (\text{tr}\{\mathbf{R}^{-1}\}\mathbf{R} - 2\eta \mathbf{R}^{-1} + \eta^2 \mathbf{R}^{-2}). \end{aligned} \quad (44)$$

Since $\bar{\mathbf{S}}$ and $(1-\lambda)^2 g$ are finite and time-invariant, (39) is stable if $\bar{\mathbf{S}}$ is stable, i.e., all eigenvalues of $\bar{\mathbf{S}}$ are smaller than one in absolute value [31] or equivalently

$$\rho\{\bar{\mathbf{S}}\} < 1. \quad (45)$$

According to A4 ($\mathbf{R}$ is assumed positive-definite) and the sub-additive inequality of the spectral radius for any two multiplication-commutative matrices [36], i.e.,

$$\rho\{\mathbf{A} + \mathbf{B}\} \leq \rho\{\mathbf{A}\} + \rho\{\mathbf{B}\},$$

where $\mathbf{AB} = \mathbf{BA}$, (45) is satisfied if

$$1 - 2\lambda + 2\lambda^2 + (1-\lambda)^2 \left( \text{tr}\{\mathbf{R}^{-1}\}\zeta_{\max}\{\mathbf{R}\} - \frac{2\eta}{\zeta_{\min}\{\mathbf{R}\}} + \frac{\eta^2}{\zeta_{\min}^2\{\mathbf{R}\}} \right) < 1.$$

Therefore, choosing any forgetting factor that satisfies

$$\lambda > 1 - \frac{2}{\text{tr}\{\mathbf{R}^{-1}\}\zeta_{\max}\{\mathbf{R}\} + \left(1 - \frac{\eta}{\zeta_{\min}\{\mathbf{R}\}}\right)^2 + 1} \quad (46)$$

guarantees mean-square stability of the RTLS algorithm.

*F. Mean-Square Deviation*

The forgetting factor, $\lambda$, is usually set close to unity. Thus, $(1-\lambda)^2$ has a small value and we may neglect the second additive term on the right-hand side of (44) to obtain

$$\bar{\mathbf{S}} \approx (1 - 2\lambda + 2\lambda^2)\mathbf{I}.$$

Subsequently, we can rewrite (39) as

$$E[\|\check{\mathbf{w}}_n\|^2] = (1 - 2\lambda + 2\lambda^2) E[\|\check{\mathbf{w}}_{n-1}\|^2] + (1-\lambda)^2 g. \quad (47)$$

The variance recursion (47) is stable as

$$0 < 1 - 2\lambda + 2\lambda^2 < 1.$$

Therefore, it converges to a steady state where we have

$$E[\|\check{\mathbf{w}}_\infty\|^2] = (1 - 2\lambda + 2\lambda^2) E[\|\check{\mathbf{w}}_\infty\|^2] + (1-\lambda)^2 g. \quad (48)$$

Substituting (43) into (48) gives the steady-state MSD of the RTLS algorithm as

$$E[\|\check{\mathbf{w}}_\infty\|^2] = \left(\frac{1-\lambda}{2\lambda}\right)\eta \\ \times \text{tr}\{\mathbf{R}^{-2}[(\|\mathbf{h}\|^2 + \gamma)(\mathbf{R} + \eta \mathbf{I}) + \eta \mathbf{h}\mathbf{h}^T]\}. \quad (49)$$

Note from (49) that when $\lambda = 1$, the steady-state MSD is zero, i.e., the algorithm is consistent. This in fact complies with the analytical findings of [14].

## VI. SIMULATIONS

Consider an EIV system identification problem where the system parameter vector has $L = 8$ entries and is chosen as

$$\mathbf{h} = [-0.019, -0.213, -0.600, +0.235, \\ +0.574, +0.377, -0.056, -0.254]^T.$$

The noiseless input vector, $\mathbf{x}_n$, is multivariate Gaussian with covariance matrix

$$\mathbf{R} = \mathbf{Q}\text{diag}\{\mathbf{f}\}\mathbf{Q}^T$$

where $\mathbf{Q} \in \mathbb{R}^{L \times L}$ is an arbitrary unitary matrix and the entries of $\mathbf{f} \in \mathbb{R}^{L \times 1}$, are drawn from a uniform distribution in the interval [0.2,1.8]. The input and output noises are also zero-mean i.i.d. multivariate Gaussian, i.e., $\mathbf{u}_n \sim \mathcal{N}(\mathbf{0}, \eta \mathbf{I})$ and $\nu_n \sim \mathcal{N}(0, \xi)$. To obtain the experimental results, we evaluate the expectations by taking the ensemble-average over $10^3$ independent simulation runs and compute the steady-state values by averaging over $10^3$ steady-state instances.

In Fig. 2, we plot the simulated time evolution of the MSD, i.e., $E[\|\check{\mathbf{w}}_n\|^2]$, for the RTLS [recursion of (11)], DCD-RTLS, kRTLS, xRTLS, and AIP algorithms. The curves are for different values of $\eta$ and $\gamma$ when $P = 10$ ($\lambda = 0.999$), $M = 16$, $H = 1$, and $N = 1$. Fig. 2 shows that the learning curves of the RTLS and DCD-RTLS algorithms overlay. This means that, even when exercising only a single iteration of the DCD algorithm at each time instant, the solutions provided by the DCD algorithm are sufficiently accurate and the assumption made at the beginning of Section V is realistic.

We observe from Fig. 2 that the DCD-RTLS algorithm converges faster than its contenders, the kRTLS, xRTLS, and AIP algorithms. Moreover, all the considered algorithms perform similarly at the steady state. Therefore, the theoretical steady-state MSD of (49) can be used to predict the steady-state MSD of all the considered algorithms. The similarity in the steady-state performance of the considered algorithms can be attributed to the fact that all these algorithms principally pursue the same objective. They calculate the adaptive FIR filter weights by recursively estimating the minor component of the augmented and weighted data covariance matrix $\boldsymbol{\Psi}_n$. The discrepancy in their transient performance is due to the different approaches taken by them to achieve the common goal. The DCD-RTLS algorithm uses the inverse power method and the DCD iterations. The kRTLS and xRTLS algorithms minimize a Rayleigh quotient cost function using the line-search method and taking steps in the direction of the Kalman gain vector or the input vector, respectively. The AIP algorithm employs the inverse power method and the line-search technique with steps being along the input vector.

In Fig. 3, we plot the lower bound on $\lambda$ given in (46) versus $\eta$ for the considered scenario where $\text{tr}\{\mathbf{R}^{-1}\} = 12.82$, $\zeta_{\min}\{\mathbf{R}\} = 0.2$, and $\zeta_{\max}\{\mathbf{R}\} = 1.8$. Notice that, to make the DCD-RTLS algorithm mean-square-stable in the presence of high input noise, $\lambda$ should be chosen close to one. However, in the presence of low input noise, mean-square stability is ensured with any $\lambda$ larger than about 0.92.

In Fig. 4, we compare the theoretical and experimental steady-state MSDs of the DCD-RTLS algorithm by plotting them against $\eta$ for different values of $\lambda$ and $\gamma$ when $M = 16$, $H = 1$, and $N = 1$. Fig. 4 shows a good match between theory and experiment.

## VII. Conclusion

We proposed a reduced-complexity recursive total least-squares (RTLS) algorithm in the context of adaptive FIR filtering employing the inverse power method and the dichotomous coordinate-descent (DCD) iterations. We utilized the DCD algorithm to solve two systems of linear equations associated with the calculation of two auxiliary vector variables. Simulation results confirmed that the proposed DCD-RTLS algorithm outperforms the other existing RTLS algorithms while being computationally more efficient. We studied the performance of the DCD-RTLS algorithm and established its convergence and stability in both mean and mean-square senses assuming that the DCD algorithm is sufficiently accurate. We also calculated the theoretical steady-state mean-square deviation (MSD) of the DCD-RTLS algorithm and verified via simulations that the theoretically predicted values of the steady-state MSD are in good agreement with the experimental results.

TABLE I
THE DCD-RTLS ALGORITHM

| | | × | + | / |
|---|---|---|---|---|
| initialize | | | | |
| $\boldsymbol{\Phi}_0 = \delta \mathbf{I}$ where $\delta$ is a small positive number | | | | |
| $\mathbf{z}_0 = \mathbf{0}, \tau_0 = 0$ | | | | |
| $\mathbf{r}_{1,0} = \mathbf{0}, \mathbf{r}_{2,0} = \mathbf{0}$ | | | | |
| $\mathbf{m}_{1,0} = \mathbf{0}, \mathbf{m}_{2,0} = \mathbf{0}$ | | | | |
| $\mathbf{w}_{-1} = \mathbf{0}, \mathbf{w}_0 = \mathbf{0}$ | | | | |
| at iteration $n = 1, 2, \dots$ | | | | |
| $\boldsymbol{\Phi}_n = \lambda \boldsymbol{\Phi}_{n-1} + \tilde{\mathbf{x}}_n \tilde{\mathbf{x}}_n^T$ | shift-structured input | $L$ | $2L$ | |
| | non-shift-structured input | $\frac{1}{2}L^2 + \frac{1}{2}L$ | $L^2 + L$ | |
| $\mathbf{z}_n = \lambda \mathbf{z}_{n-1} + \tilde{y}_n \tilde{\mathbf{x}}_n$ | | $L$ | $2L$ | |
| $\tau_n = \lambda \tau_{n-1} + \tilde{y}_n^2$ | | 1 | 2 | |
| $\mathbf{p}_{1,n} = \lambda \mathbf{r}_{1,n-1} + (\tilde{y}_n - \tilde{\mathbf{x}}_n^T \mathbf{m}_{1,n-1}) \tilde{\mathbf{x}}_n$ | | $2L$ | $3L$ | |
| $\mathbf{p}_{2,n} = \lambda (\mathbf{r}_{2,n-1} - \mathbf{w}_{n-2}) + \mathbf{w}_{n-1} - (\tilde{\mathbf{x}}_n^T \mathbf{m}_{2,n-1}) \tilde{\mathbf{x}}_n$ | | $2L$ | $5L - 1$ | |
| for $i = 1, 2$ | | | | |
| solve $\boldsymbol{\Phi}_n \mathbf{d}_{i,n} = \mathbf{p}_{i,n}$ to obtain $\mathbf{d}_{i,n}$ and $\mathbf{r}_{i,n}$ | | 0 | $2NL + N + M$ | |
| $\mathbf{m}_{i,n} = \mathbf{m}_{i,n-1} + \mathbf{d}_{i,n}$ | | 0 | $L$ | |
| $\mathbf{k}_n = \mathbf{m}_{1,n} + \gamma^{-1} \tau_n \mathbf{m}_{2,n}$ | | $L + 1$ | $L$ | |
| $\mathbf{w}_n = \mathbf{k}_n - \frac{\mathbf{z}_n^T \mathbf{k}_n}{\gamma + \mathbf{z}_n^T \mathbf{m}_{2,n}} \mathbf{m}_{2,n}$ | | $3L$ | $2L - 1$ | 1 |

TABLE II
THE DCD ALGORITHM SOLVING $\boldsymbol{\Phi}_n \mathbf{d}_{i,n} = \mathbf{p}_{i,n}$

| | + |
|---|---|
| initialize | |
| $\epsilon = 1, \alpha = H/2$ | |
| $\mathbf{d}_{i,n} = \mathbf{0}$ | |
| $\mathbf{r}_{i,n} = \mathbf{p}_{i,n}$ | |
| for $j = 1, 2, \dots, N$ | |
| $l = \arg\max_{k=1,\dots,L}\{|r_{k,i,n}|\}$ | $L - 1$ |
| while $|r_{l,i,n}| \leq \frac{\alpha}{2} \phi_{l,l,n}$ and $\epsilon \leq M$ | 1 |
| $\epsilon = \epsilon + 1, \alpha = \alpha/2$ | |
| if $\epsilon > M$ | |
| algorithm stops | |
| $d_{l,i,n} = d_{l,i,n} + \text{sign}\{r_{l,i,n}\}\alpha$ | 1 |
| $\mathbf{r}_{i,n} = \mathbf{r}_{i,n} - \text{sign}\{r_{l,i,n}\}\alpha \boldsymbol{\phi}_{l,n}$ | $L$ |

TABLE III
COMPUTATIONAL COMPLEXITY OF DIFFERENT ALGORITHMS IN TERMS OF
NUMBER OF REQUIRED ARITHMETIC OPERATIONS PER ITERATION

| | × | + | / | √ |
|---|---|---|---|---|
| | shift-structured input | | | |
| DCD-RTLS | $10L + 2$ | $(4N + 17)L + 2N + 2M$ | 1 | 0 |
| AIP | $15L + 11$ | $12L + 5$ | 1 | 0 |
| xRTLS | $16L + 19$ | $13L + 5$ | 2 | 1 |
| kRTLS | $22L + 93$ | $19L + 47$ | 8 | 2 |
| | non-shift-structured input | | | |
| DCD-RTLS | $0.5L^2 + 9.5L + 2$ | $L^2 + (4N + 16)L + 2N + 2M$ | 1 | 0 |
| AIP | $2L^2 + 9L + 9$ | $1.5L^2 + 6.5L + 5$ | 1 | 0 |
| xRTLS | $2L^2 + 10L + 17$ | $1.5L^2 + 7.5L + 5$ | 2 | 1 |
| kRTLS | $3L^2 + 10L + 31$ | $2L^2 + 6L + 13$ | 6 | 2 |

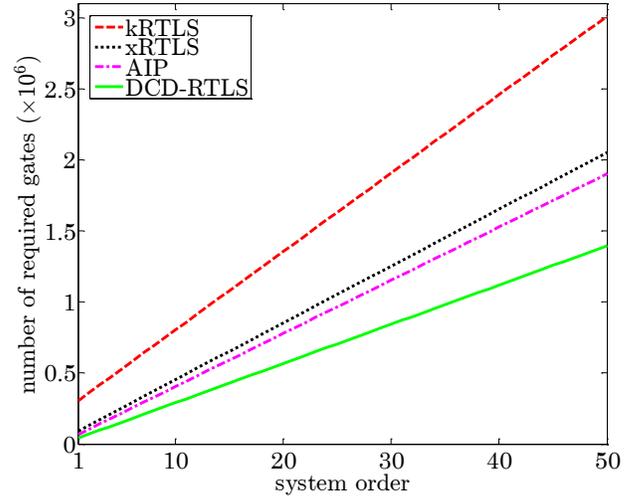

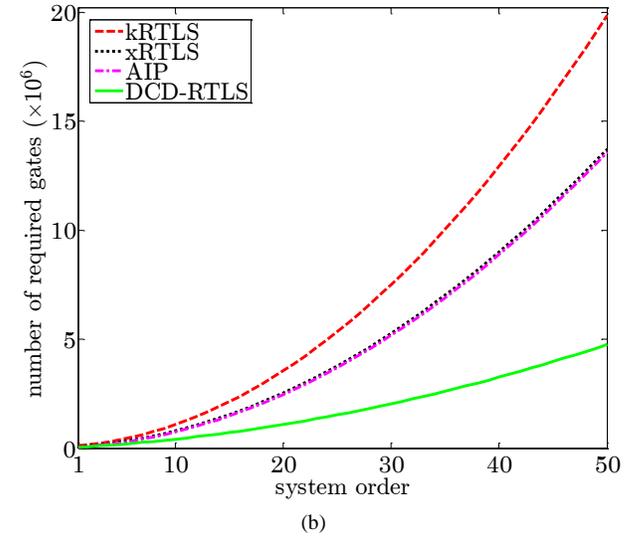

Fig. 1. Number of required gates by different algorithms when the input data is (a) shift-structured or (b) non-shift-structured. In the DCD-RTLS algorithm, $M = 16$ and $N = 1$.

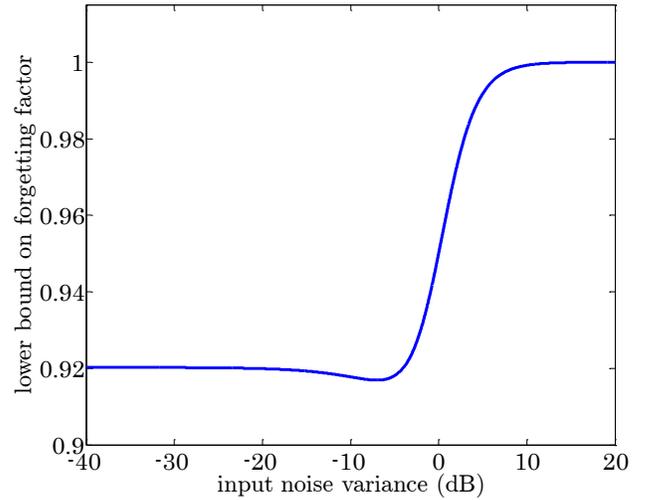

Fig. 3. The lower bound on the forgetting factor for ensuring mean-square stability of the DCD-RTLS algorithm as a function of the input noise variance.

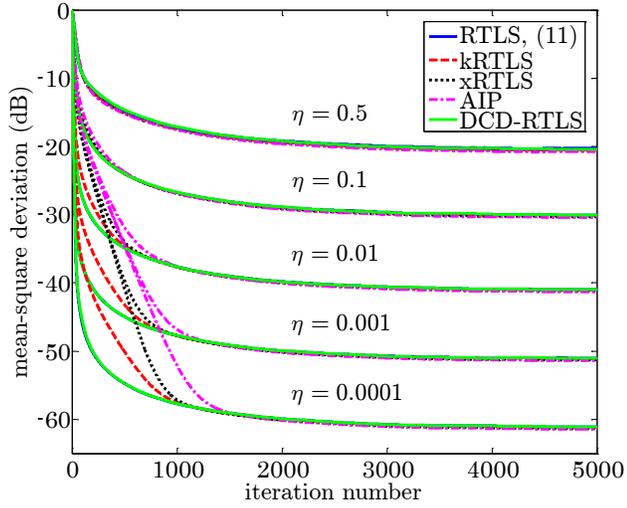
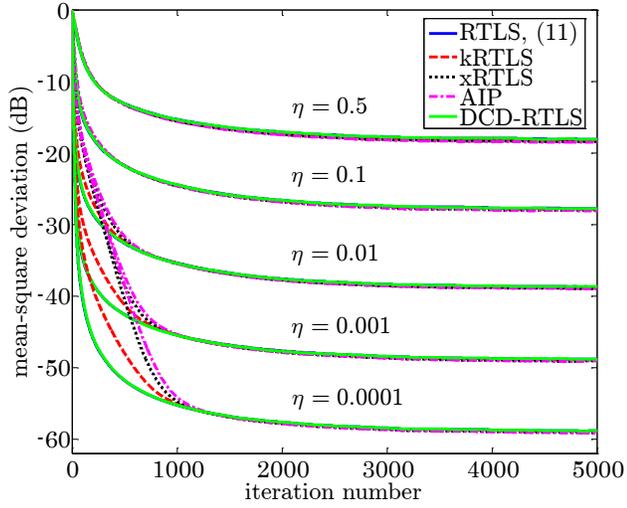
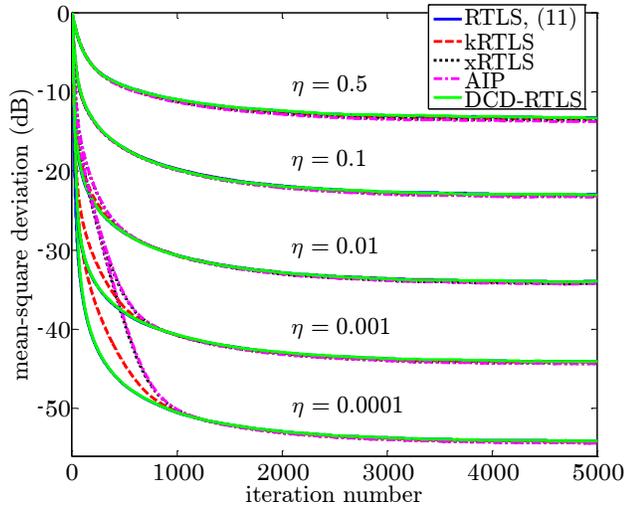

Fig. 2. Learning curves of different algorithms for different values of the input noise variance when $P = 10$ ($\lambda = 0.999$), $M = 16$, $H = 1$, $N = 1$, and (a) $\gamma = 1/5$, (b) $\gamma = 1$, or (c) $\gamma = 5$.

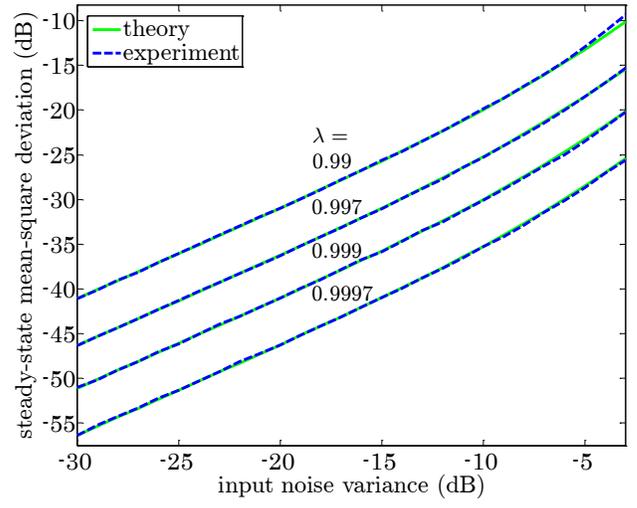
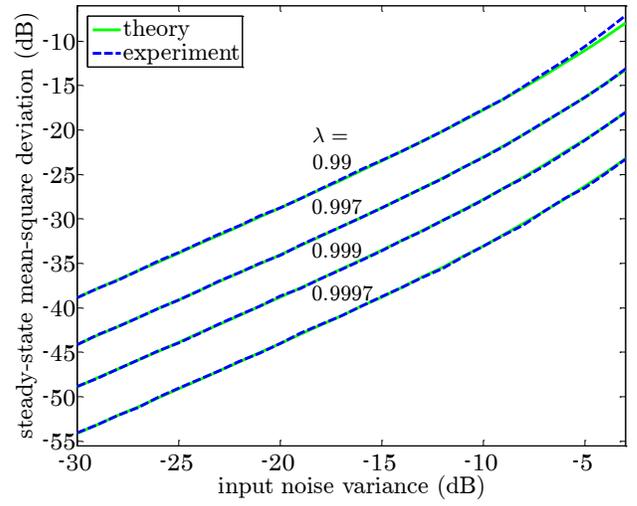
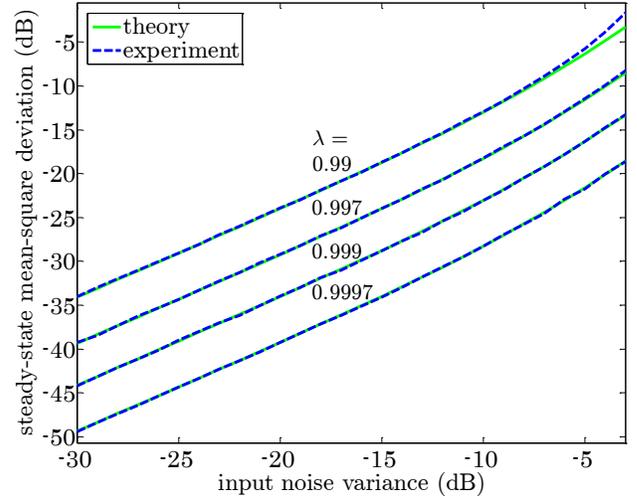

Fig. 4. Theoretical and experimental steady-state MSDs of the DCD-RTLS algorithm as a function of the input noise variance for different values of the forgetting factor when $M = 16$, $H = 1$, $N = 1$ and (a) $\gamma = 1/5$, (b) $\gamma = 1$, or (c) $\gamma = 5$.